# An antisymmetric plasmon resonance in coupled gold nanoparticles as a sensitive tool for detection of local index of refraction.

A. N. Grigorenko*, H. F. Gleeson, Y. Zhang, N. Roberts,

*Department of Physics and Astronomy, University of Manchester, Manchester, M13 9PL, UK*

A. R. Sidorov,

*General Physics Institute, 38 Vavilov str., Moscow 117942, Russia*

A. A. Panteleev,

*Troitsk Institute of Innovation and Thermonuclear Investigations, Troitsk 142190, Russia*

**Abstract:**

A nanofabricated regular array of coupled gold nano-pillars is employed to detect local indices of refraction of different liquids using a shift of an antisymmetric plasmon resonance peak observed in the reflection spectra. The peak's spectral position is found to be a unique function of the local refractive index for a wide range of indices. We discuss possible applications of the fabricated nanostructured arrays in bio- and chemical sensors.

* corresponding author: sasha@man.ac.uk

Plasmon resonances (selective excitation of localised vibrations of the electron plasma in metals by electromagnetic field) turned out to be a valuable tool in the development of new optical devices. Surface plasmons [1] produced in different configurations (e.g., in Kretschmann [2] or Otto [3] geometries) have been extensively used in a large variety of tools and techniques for bio- and chemical sensing, such as surface plasmon resonance (SPR) sensors [4], commercially available from Biacore, surface plasmon microscopy [5], colorimetric detection of DNA by gold nanoparticles [6], etc. The sensitivity of surface plasmon techniques has been improved to just one atomic monolayer by using interferometry [7, 8] and dark-field techniques [9].

Most of the initial success of plasmonic optics was achieved by exploiting plasmon resonances in thin metallic films. However, the advent of nanofabrication allowed one to create controlled arrays of nanoparticles and nanostructures, also possessing plasmon resonant modes of electron vibrations. These plasmon modes retain the high sensitivity of plasmon resonances to the environment but provide a local response and compatibility with the high-throughput techniques required by biophysics. As a result, plasmonic optics has benefited from the discovery of plasmon assisted extraordinary transmission of light [10], detection of polynucleotides based on distance-dependent optical properties of nanoparticles [11], light concentration in self-similar chains of metal nanospheres [12], detection of local refractive index from individual nanoparticles [13]. Nanoparticles with resonant plasmonic modes made it possible to create novel artificial materials based on metallic "nano-atoms" and "nano-molecules" with unique optical properties unachievable in natural materials.

Recently we have fabricated a new artificial nano-medium formed by regular arrays of "nano-molecules" - pairs of coupled identical gold nano-pillars with plasmon resonances in the visible part of the spectrum [14]. In this material, which may be referred to as optomagnetic [14, 15], a plasmon resonance mode of a nanoparticle splits into two resonant modes for a particle pair: a symmetric mode, characterized by an overall dipole response and contributing to the medium's permittivity, and an antisymmetric mode, in which the dipoles oscillate in antiphase cancelling the dipole response and leaving the overall magnetic response contributing to the magnetic permeability [15,16]. The purpose of this work is to show that the antisymmetric plasmon resonance mode of the coupled nano-pillars can be used to detect local changes of the refractive index with high sensitivity in a wide range of refractive indices. Combined with a relatively simple design, controlled nature of resonance splitting, compatibility with high-throughput screening and a simple optical scheme for plasmon excitation, the proposed medium has a potential to find a wide range of applications in bio- and chemical sensors.

Figure 1(a) shows an electron micrograph of one of our samples. The prepared structures were regular arrays of Au pillars fabricated by high-resolution electron-beam lithography on a glass substrate and grouped in tightly spaced pairs. The structures typically covered an area of $\approx 0.1 \text{mm}^2$ and contained $\approx 10^6$ pillars. The lattice constant, $a$, for periodic arrays was in the range 400-600nm. Heights $h$ of Au pillars and their diameters were chosen through numerical simulations so that the plasmon resonance in the reference samples appeared at red-light wavelengths. A number of different structures were studied with $d$ between 80 and 160nm and the pair separations $s$ between the centres of adjacent pillars in the range 140 to 200nm and, i.e. the gap $s$–$d$ between the neighbouring pillars varied from 100nm down to almost zero (overlapping pillars). At these separations, the electromagnetic interaction between the nanopillars splits plasmon resonances, and plasmonic modes of the

"nano-molecule" can be characterised by their parity. Figures 1(b) and (c) show the current distributions for the symmetric and antisymmetric $z$-modes, respectively, calculated with the electromagnetic module of Femlab software for the actual experimental geometry. The symmetric $z$-mode is characterised by non-zero dipole moment directed along the $z$-axis. The overall dipole moment of the antisymmetric $z$-mode is zero and the circulating currents in the $x$-$z$ plane produce the magnetic moment along the $y$-axis. There exist three main symmetric and three antisymmetric resonant modes in an interacting pair with currents flowing along the $x$-, $y$- and $z$-axes. Excitation of these modes depends on the symmetry of the pillar pair and the conditions of light incidence. In our case, an efficient coupling of resonant modes to incident light has been insured by non-cylindrical geometry of pillars which was intentionally introduced in the design through a choice of microfabrication procedures (double layered resist).

Figures 1(d) and (e) show typical reflection spectra measured from the sample of Fig. 1(a) under the condition of normal light incidence for TM light (with the electric field vector along the $x$-axis) and TE light (with the electric field vector along the $y$-axis), respectively. The reflection spectra were collected by a conventional spectrophotometer from an area of $100 \times 100$ μm$^2$. There are two distinct resonance peaks in the TM spectrum indicated by arrows and only one peak in the TE spectrum. The symmetry analysis and the numerical solution of Maxwell equations for the experimental geometry with Femlab software [14] (the results are shown in the insets of Fig. 1(d) and (e) for the resonant wavelengths) proves that the weaker resonance peak (observed at green wavelength) corresponds to the antisymmetric $z$-mode and the stronger peaks of Fig. 1(d) and (e) (observed at red wavelengths) correspond to the symmetric $x$- and $y$-modes of plasmonic resonances, respectively. It is worth noting that diffraction modes do not contribute to generation of the resonant peaks, as the fabricated structures with different periodicities ($a$=400nm and 600nm) and random arrays of pairs of the same geometry (sizes and separation inside the pair) have resonance peaks at the same spectral positions.

The plasmonic resonances shown in Figs. 1(d) and (e) are excited by light of normal incidence (see schematic view of the experimental set-up in Fig. 2(a)) and do not require an elaborate optical scheme of Kretschmann or Otto geometries. This implies that our medium can be easily combined with sensing techniques based on any kind of reflection or microscopy. We demonstrate this by employing our structures to detect local refractive indices of optically-thin liquid layers. Figure 2(b) shows the TM spectra of the sample of Fig. 1(a) covered with a thin layer of liquid taken at intervals as the liquid was drying (curves 1-3). Initially, the reflection spectrum (curve 1) shows an interference pattern produced by the interference of light reflected from the top and the bottom of the liquid layer. The positions of minima in the interference pattern can be found from the condition [17] $nD = (2m+1)\lambda/4$, where $n$ is the refractive index of the liquid layer, $D$ is the layer thickness and $m \geq 0$ is an integer. As the liquid dries up, $D$ decreases and the interference pattern gradually disappears (compare curves 1 and 2) eventually leaving spectrum 3 without any trace of interference (which implies that $D<<\lambda/4n$). The final spectrum (curve 3) shows the peaks of the plasmon resonances due to the nanostructured media and did not change for several hours after initial drying (even days for some liquids). Using Fresnel reflection coefficients [17] to model the change of the interference pattern of Fig. 2(b), we can obtain a rough estimate of the average effective thickness of dried liquid as <50nm. We believe, however, that the liquid is mostly concentrated "inside" the "nano-molecules" (between pillars in a pair) and is dried completely in areas between

"nano-molecules". Below we shall refer to the final stable spectrum as the spectrum of samples covered with optically-thin layers of liquid. Before every series of measurements, the samples were cleaned by rinsing in acetone, propanol and blowing dry by compressed nitrogen. The cleaning procedure removed the liquid completely and restored the original reflection spectrum of the sample shown in Fig. 1(d) and (e).

The important property of the spectra of structures covered with an optically-thin layer of liquid is an increase of the amplitude of the antisymmetric resonance and a substantial red shift of the resonance peak. (It worth noting that the symmetric resonance peak was also red-shifted albeit decreased in the amplitude [14].) These features are emphasised in Fig. 3(a), where the spectra measured in normal TM light for the sample of Fig. 1(a) covered with different liquids are presented. One can notice that the change in reflection around the antisymmetric resonance for, e. g., spectrum 4 of Fig. 3(a) is ≈6% (the spectrum corresponds to optical coupling fluid OCF463). This change is about 3 times greater than that for the spectrum of the clean sample (Fig. 1(b), ≈2%). It is also clear that the resonance peak is red-shifted in Fig. 3(a) by ≈30-60nm with respect to the original spectral peak's position in TM spectrum of Fig. 1(b). We extracted the spectral position of the antisymmetric resonance peak from the spectra and plotted it as a function of the refractive index of the liquid in Fig. 3(b). Figure 3(b) demonstrates the main experimental result of this letter: in a wide range of refractive indices (from 1.3 to 1.7) there exists a unique (close to linear) dependence between the refractive index of an optically-thin liquid layer covering the fabricated metamaterial and the shift of the antisymmetric resonance. The observed shift was about 2.8nm per 0.01 index of refraction. This is comparable with the red-shifts recorded previously for the quadrupole resonance of an individual nanoparticle [13] (1.6nm per 0.01) but less than that for the conventional

SPR technique (≈5-7nm per 0.01) [1,2]. However, none of the previous techniques works in such a large range of refractive indices and has the simplicity of the plasmon excitation and spectrum registration.

We briefly discuss the dependence of the resonance shifts on the refractive index of the environment. The reflection spectra of our samples can be calculated numerically by direct solving of Maxwell equations for the experimental geometry, see Ref. [14] for details. The extracted resonance shift is a complex function of the refractive index of the environment (determined by the structure geometry, metal constants and substrate properties). It would be of great benefit, however, to approximate this function by a simple analytical expression. This cannot be done within the well-developed Mie theory because of non-spherical nature of nano-pillars and the strong electromagnetic interaction between pillars in the pairs. Instead, we use a theory [15], which takes into account the effects of the particle interaction and the environment through Leontovich boundary conditions and yields an analytical expression for the resonance wavelengths of a double-wire nano-molecule. The wavelength of the antisymmetric resonance $\lambda_{anti}$ can be found from the implicit formula [15]

$$\lambda = 2h \operatorname{Re}\left[\sqrt{\varepsilon\left(1 + \frac{i\lambda\varsigma_{xx}}{4\pi d \ \ln(2s/d)}\right)}\right], \qquad (1)$$

where $\varepsilon$ is the local permittivity, $c$ the speed of light and $\varsigma_{xx} = \varsigma_{xx}(\lambda)$ is the element of the surface impedance matrix. For the strong normal skin-effect we can simplify (1) to

$$\lambda_{anti} \approx 2nh\sqrt{1 + \frac{c\sqrt{\tau/(4\pi\sigma_0)}}{\pi d \ \ln(2s/d)}} = \lambda_r \cdot n, \qquad (2)$$

where $\sigma_0 = \omega_p^2 \tau/4\pi$, $\omega_p$ is the plasma frequency, $\tau$ is the relaxation time, $\lambda_r$ is a constant. Equation (2) suggests a linear dependence of the resonance wavelength on the local refractive index $n$. For the intermediate skin-effect, which is applicable to our structures, (1) is simplified to $\lambda \approx \lambda_r n(1-\lambda/\Lambda)$, giving the resonance wavelength as

$$\lambda_{anti} \approx \lambda_r \frac{n}{1+a \cdot n}, \qquad (3)$$

where $a = \lambda_r / \Lambda$ and $\Lambda$ is a constant. The simple expression (3) describes well the measured dependence presented in Fig. 3(b). It is necessary to stress that (3) is calculated near the limit of the applicability of the theory [15]. (The theory requires $2h>d$ and works the best at $2h>>d$ while $2h/d$ is $\approx 1.2$ in our case.) For this reason, we regard (1)-(3) only as a useful analytical approximation which is proved to work reasonably well for the studied structures.

The lateral resolution with which the reflection spectra were recorded in our experiments was about 100μm. It can be reduced to 10μm by using stronger objective lenses and/or by engineering microarrays of cells (based on the proposed metastructures) either with electron lithography or microcontact printing [18]. This would allow one to study the refractive index of an unknown liquid and to perform high-throughput screening [19] on an almost micrometer scale. The spatial resolution of the analysis can be improved further down to 1μm by using other methods of surface microscopies. For example, it is known that electromagnetic fields are greatly enhanced near the metal in the presence of plasmonic modes, which leads to surface enhanced Raman scattering (SERS) [20, 21]. This implies that the fabricated nano-medium might be useful as a substrate for SERS. Indeed, Fig. 4 shows a fragment of the Raman spectra measured using Renishaw Raman Microscope (System 2000) for the structures of Fig. 1(a) (illuminated by TM and TE light of wavelength $\lambda$=532nm) and the glass substrate covered by 2-propanol. In all measurements, the SERS spectra were collected from a 1μm spot (using ×50 objective). The magnitude of the SERS peak for the structure is about 4000 times larger than for the glass substrate, which gives a rough estimate of the local field enhancement of 8, confirmed by the Femlab numerical calculations described above. We also note that the SERS peak for TM polarization (where the antisymmetric mode of the plasmon resonance is excited at $\lambda$=532nm) is about 7 times greater than for TE polarization, in agreement with the reflection spectra of Fig. 1(b). It is apparent that SERS is superior in resolution and sensitivity, as compared to measurements of the reflection spectra. However, the practicality of its use for cheap high-throughput screening is doubtful.

In conclusion, we have demonstrated that the shift of the antisymmetric plasmon resonance of the media produced by pairs of identical gold nano-pillars can be applied to detect local refractive indices of optically-thin layers of liquids. This method provides a micrometer spatial resolution and a very high sensitivity ($\approx$3nm per 0.01 of the refractive index) in a wide range of refractive indices (from 1.3 to 1.7). The simple optical scheme for plasmon excitation and compatibility with high-throughput analysis suggest that the fabricated media can be successfully used for bio- and chemical sensing.

Acknowledgements: Authors thank L. Panina and D. Makhnovskiy for fruitful discussions. The help of Research Support Fund of the University of Manchester is acknowledged.

Figure Captions.

Fig. 1. Fabricated optomagnetic medium and its spectra. (a) A micrograph of the sample ($a$=500nm, $s$=200nm and $d$=160nm) on which all data presented were collected. (b) The distribution of electric currents (conical arrows) inside a pair of pillars for the resonant symmetric $z$-mode. (c) Same for the antisymmetric $z$-mode. (d, e) Experimental reflection spectra measured for TM and TE polarizations, respectively. The insets show the current distribution calculated by solving Maxwell equations for the actual experimental geometry at the resonant wavelengths.

Fig. 2. (a) Optical scheme for measurements of the reflection spectra. (b) The dynamics of the reflection spectra of the sample covered with a thin layer of glycerol ($n$=1.47): 1 – the initial spectrum; 2 – the spectrum after 15 minutes; 3 –after 1 hour. The spectra are offset for clarity. Drying was assisted by a fan and illumination.

Fig. 3. (a) Reflection spectra registered in the presence of optically-thin layers of liquids: 1 – clean sample; 2 – glycerol+water; 3 – glycerol; 4 – OCF463. (b) The antisymmetric resonance wavelength as a function of the refractive index of an optically-thin liquid layer.

Fig. 4. Fragment of the Raman spectra measured for the sample and the glass substrate covered with a thin layer of 2-propanol in TM and TE polarizations. The spectra are offset for clarity.

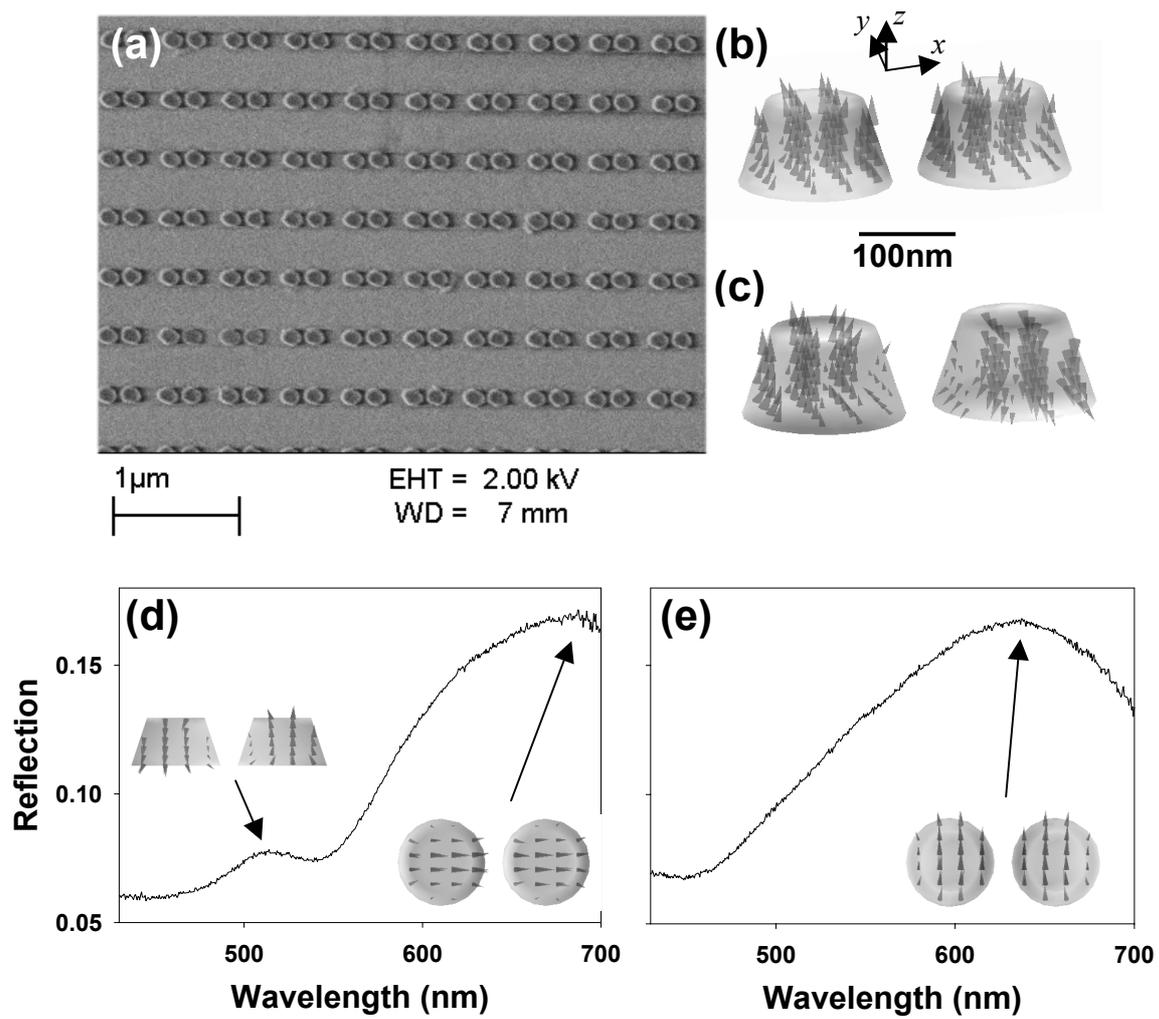

Fig. 1.

*Grigorenko et al.*

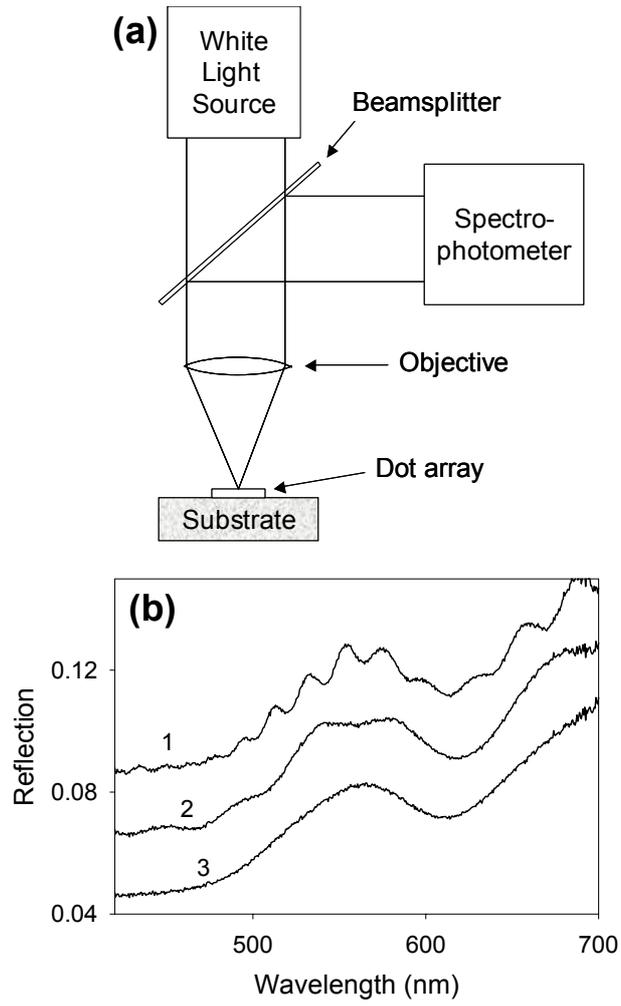

Fig. 2.

*Grigorenko et al.*

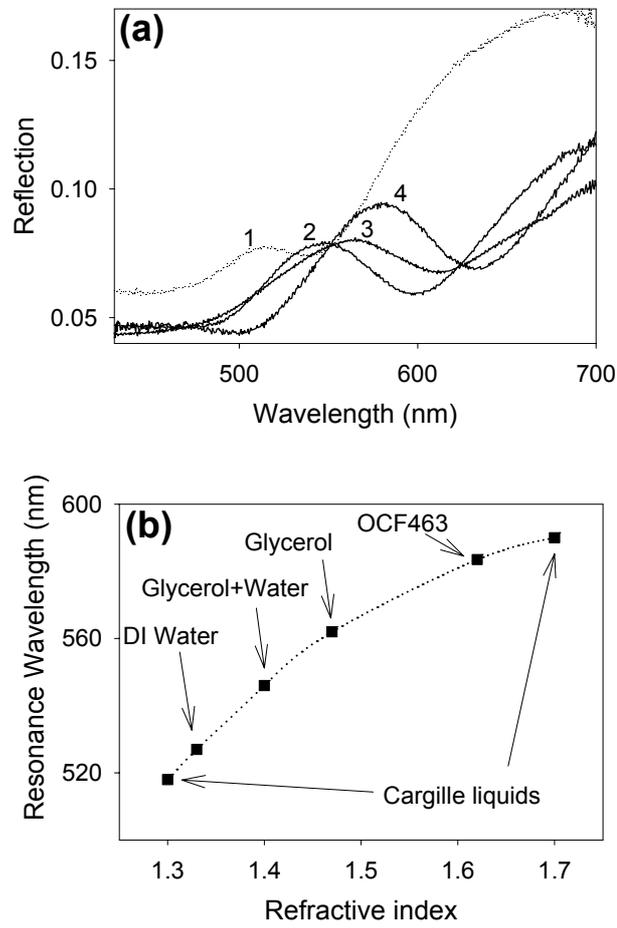

Fig. 3.

*Grigorenko et al.*

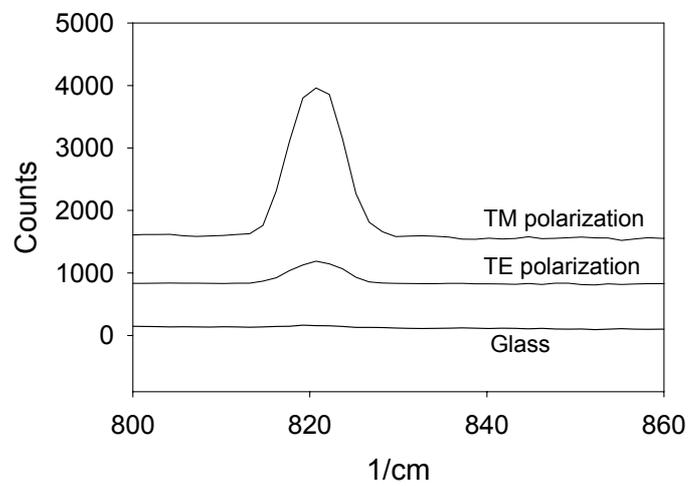

Fig. 4.

*Grigorenko et al.*